\documentclass[12pt]{article}
\usepackage{amsfonts}
\usepackage[dvips]{graphicx}
\usepackage{latexsym,amsmath,amssymb,graphics,stmaryrd}
\usepackage{cite}
\usepackage{array}
\usepackage{subfigure}

\usepackage{multirow}
\usepackage{epsfig}
\usepackage[dvips]{graphicx}
\usepackage[dvips]{color}
\usepackage{pstricks}
\usepackage{pst-node}
\usepackage{pst-plot}
\usepackage{dsfont}
\usepackage{amsthm}
\usepackage{graphics}
\usepackage{subfigure}
\usepackage{fancyhdr}
\usepackage[dvips]{color}

\usepackage{feynmp}
\fancypagestyle{plain}{
\fancyhf{}
\newcommand{\fpage}{\iffloatpage{}{\thepage}}
\fancyfoot[C]{\fpage}

}

\newcommand{\col}{~,}
\newcommand{\pnt}{~.}
\newcommand{\AdS}{\text{AdS}}
\newcommand{\CFT}{\text{CFT}}
\newcommand{\YM}{\text{YM}}

\newcommand{\comm}[2]{\left[#1\smash[b]{\mathbin{,}}#2\right]}

\newcommand{\de}{\operatorname{d}\!}

\newcommand{\e}{\operatorname{e}}
\newcommand{\pfour}[4]{{}\{#1,#2,#3,#4\}{}}
\newcommand{\pthree}[3]{{}\{#1,#2,#3\}{}}
\newcommand{\ptwo}[2]{{}\{#1,#2\}{}}
\newcommand{\pone}[1]{{}\{#1\}{}}
\newcommand{\pid}{{}\{ \}{}}

\newlength{\neglength}
\newlength{\diameter}
    
\DeclareMathOperator{\tr}{tr}

\DeclareMathOperator{\perm}{P}
\numberwithin{equation}{section}
\newlength{\eqoff}
\newlength{\eqofftwo}
\newlength{\unit}
\psset{xunit=\unit,yunit=\unit,runit=\unit}
\newlength{\linew}
\psset{linewidth=\linew}
\begin{document}
\begin{fmffile}{lettgraphs}

\fmfcmd{%
marksize=1.5mm;
def draw_mark(expr p,a) =
  begingroup
    save t,tip,dma,dmb; pair tip,dma,dmb;
    t=arctime a of p;
    tip =marksize*unitvector direction t of p;
    dma =marksize*unitvector direction t of p rotated -45;
    dmb =marksize*unitvector direction t of p rotated 45;
    linejoin:=beveled;
    draw (-.5dma.. .5tip-- -.5dmb) shifted point t of p;
  endgroup
enddef;
style_def derplain expr p =
    save amid;
    amid=.5*arclength p;
    draw_mark(p, amid);
    draw p;
enddef;
def draw_point(expr p,a) =
  begingroup
    save t,tip,dma,dmb,dmo; pair tip,dma,dmb,dmo;
    t=arctime a of p;
    tip =marksize*unitvector direction t of p;
    dma =marksize*unitvector direction t of p rotated -45;
    dmb =marksize*unitvector direction t of p rotated 45;
    dmo =marksize*unitvector direction t of p rotated 90;
    linejoin:=beveled;
    draw (-.5dma.. .5tip-- -.5dmb) shifted point t of p withcolor 0white;
    draw (-.5dmo.. .5dmo) shifted point t of p;
  endgroup
enddef;
style_def derplainpt expr p =
    save amid;
    amid=.5*arclength p;
    draw_point(p, amid);
    draw p;
enddef;
style_def dblderplain expr p =
    save amidm;
    save amidp;
    amidm=.5*arclength p-0.5mm;
    amidp=.5*arclength p+0.5mm;
    draw_mark(p, amidm);
    draw_point(p, amidp);
    draw p;
enddef;
}

\begin{titlepage}
\begin{flushright}
IFUM-910-FT\\
\end{flushright}
\vspace{5ex}
\Large
\begin {center}     
{\bf
Wrapping at four loops in $\mathcal{N}=4$ SYM}
\end {center}

\renewcommand{\thefootnote}{\fnsymbol{footnote}}

\large
\vspace{1cm}
\centerline{F.\ Fiamberti ${}^{a,b}$, A.\ Santambrogio ${}^b$, 
C.\ Sieg ${}^b$, D.\ Zanon ${}^{a,b}$
\footnote[1]{\noindent \tt
francesco.fiamberti@mi.infn.it \\
\hspace*{6.3mm}alberto.santambrogio@mi.infn.it \\ 
\hspace*{6.3mm}csieg@mi.infn.it \\ 
\hspace*{6.3mm}daniela.zanon@mi.infn.it}}
\vspace{4ex}
\normalsize
\begin{center}
\emph{$^a$  Dipartimento di Fisica, Universit\`a degli Studi di Milano, \\
Via Celoria 16, 20133 Milano, Italy}\\
\vspace{0.2cm}
\emph{$^b$ INFN--Sezione di Milano,\\
Via Celoria 16, 20133 Milano, Italy}
\end{center}
\vspace{0.5cm}
\rm
\abstract
\normalsize 
We present the planar four-loop anomalous dimension of the composite operator
$\tr(\phi\comm{Z}{\phi} Z)$ in the flavour $SU(2)$ sector of the
${\cal{N}}=4$ SYM theory. At this loop order wrapping interactions are
present: they give rise to contributions proportional to $\zeta(5)$
increasing the level of transcendentality of the anomalous dimension. In
a sequel of this paper all the details of our calculation will be
reported.

\vfill
\end{titlepage} 

\section{Introduction}

After the advent of the $\AdS/\CFT$ conjecture \cite{maldacena:1998re} there has 
been a
renewed interest in $\mathcal{N}=4$ SYM theory, which represents one
of the best playgrounds to test new ideas connected to non-perturbative
results.

The first prediction that one would like to test
is the matching of the spectra on the two sides  of the conjecture. In fact we expect 
that the spectrum of the anomalous dimensions of gauge invariant
operators of the planar $\mathcal{N}=4$ SYM theory matches the spectrum of strings
on $\AdS_5\times\text{S}^5$.

Thus it is very important to have tools for the computation of
anomalous dimensions. A big progress in this direction has been
made in the last five years after the realization \cite{Minahan:2002ve,
Beisert:2003tq} that
the planar one-loop dilatation operator of $\mathcal{N}=4$ SYM maps into the
Hamiltonian of an integrable spin chain. The spin chain picture
revealed itself very fruitful in understanding the
integrability properties of higher orders in perturbation
theory \cite{Beisert:2003ys,Staudacher:2004tk}. In addition it suggested to compute anomalous dimensions while finding solutions of associated Bethe equations \cite{Beisert:2004hm}. 
The form of these equations has been recently refined with the introduction 
of the so-called dressing phase \cite{Arutyunov:2004vx,Hernandez:2006tk,Beisert:2006ib,
Beisert:2006ez}.

Now the hope of a direct
comparison between the spectra on the two sides of the $\AdS/\CFT$
correspondence is definitely more concrete.
However a major obstacle in pursuing this program is due to the fact that
the Bethe ansatz is  asymptotic, i.e.
 it applies only to long operators. Indeed the
 spin chain Hamiltonian is long range: at a given
perturbative order $K$ in the coupling constant
\begin{equation}\label{gdef}
g=\frac{\sqrt{\lambda}}{4\pi}
\end{equation}
(where $\lambda=g^2_\YM N$ is the 't Hooft coupling) the range of the 
interactions between
adjacent sites grows with the perturbative order as $K+1$. For an
operator of length $L$ we should expect new effects when the range
exceeds $L$.
The asymptotic Bethe ansatz  breaks down at orders $K\ge L$
since the interaction is 
no longer localized in some limited region along the state and asymptotic 
states cannot  be defined. 
This spreading of the interaction manifests itself with the insurgence of a new type of
 contributions, 
 the so-called {\it wrapping interactions}. 

Several papers have addressed this issue.
In \cite{Sieg:2005kd} the properties of wrapping interactions
have been worked out in terms of Feynman diagrams. 
In \cite{Ambjorn:2005wa} it was proposed that 
the thermodynamic Bethe ansatz captures the wrapping interactions. 
This work has been extended in \cite{Arutyunov:2006gs,Janik:2007wt}.
In \cite{Rej:2005qt} it was assumed that wrapping might be described by 
the Hubbard model. 
Finally by using restrictions from the BFKL equation 
\cite{Lipatov:1976zz,Kuraev:1977fs,Balitsky:1978ic},
a quantitative proposal for the anomalous dimensions at four-loop order with
 wrapping effects was conjectured in \cite{Kotikov:2007cy}.

The aim of this Letter is to shed light on the present situation: we perform
an explicit field theoretical anomalous dimension computation 
for the length-four
Konishi descendant $\tr(\phi\comm{Z}{\phi}Z)$ at four loops. This is the 
simplest case
in which wrapping effects are present.
Different conjectures on the value of this anomalous dimension were proposed in
\cite{Rej:2005qt,Beisert:2006ez,Kotikov:2007cy}.

A complete four-loop Feynman graph calculation is terribly complicated, so  
we had
to find a simplifying strategy in order to proceed. The diagrams are naturally 
divided in two classes:
four-loop graphs with no wrapping and graphs with wrapping interactions. We 
obtain the contributions from the two classes as follows:

First we take advantage of the known form of the four-loop dilatation 
operator $D_4$ given in \cite{Beisert:2007hz}: all the non-wrapping 
contributions can be 
obtained by subtracting from $D_4$ its range 5 part. 
The remaining terms contain all the contributions with range 
from 1 to 4, so they can be applied safely to our length four operator. 
In this way we avoid the explicit computation of this vast and difficult 
class of Feynman graphs. This will be done in Section 2.

We consider wrapping interactions in Section 3. Many diagrams need to be
considered but $\mathcal{N}=1$ supergraph techniques allow us to drastically 
simplify the calculation. After  $D$-algebra manipulations the 
diagrams are reduced to standard four-loop momentum integrals which we
compute by means of uniqueness and the Gegenbauer polynomial
$x$-space technique 
\cite{Chetyrkin:1980pr,Kazakov:1983ns}.

Finally in Section 4 we collect all the terms and compute the four-loop 
planar anomalous dimension 
of the length four Konishi descendant. Our result shows that previous 
conjectures do not reproduce the correct anomalous dimension. In particular 
we find that at this loop order wrapping interactions give rise to 
contributions proportional to $\zeta(5)$
increasing the level of transcendentality of the anomalous dimension.  In the 
following we describe the various steps that allowed us to reach the final 
result, while the details of the calculation will be reported in a separate 
publication \cite{Fiamberti:2008}.

\section{Subtraction of range five interactions}

In this section we compute the contributions to the anomalous dimension due 
to four-loop non-wrapping graphs. To this end we consider   
the four-loop planar dilatation operator in the $SU(2)$ subsector containing 
all operators made of two out of the three complex scalars of $\mathcal{N}=4$
SYM, denoted by $\phi$ and $Z$.
It is given by \cite{Beisert:2007hz}

\begin{equation}\label{D4}
\begin{aligned}
D_4&=-(560+4\beta)\pid\\
&\phantom{{}={}}+(1072+12\beta+8\epsilon_{3a})\pone1\\
&\phantom{{}={}}-(84+6\beta+4\epsilon_{3a})\ptwo13
-4\ptwo14-(302+4\beta+8\epsilon_{3a})(\ptwo12+\ptwo21)\\
&\phantom{{}={}}
+(4\beta+4\epsilon_{3a}+2i\epsilon_{3c}-4i\epsilon_{3d})\pthree132
+(4\beta+4\epsilon_{3a}-2i\epsilon_{3c}+4i\epsilon_{3d})\pthree213\\
&\phantom{{}={}}+(4-2i\epsilon_{3c})(\pthree124+\pthree143)
+(4+2i\epsilon_{3c})(\pthree134+\pthree214)\\
&\phantom{{}={}}+(96+4\epsilon_{3a})(\pthree123+\pthree321)\\
&\phantom{{}={}}-(12+2\beta+4\epsilon_{3a})\pfour2132
+(18+4\epsilon_{3a})(\pfour1324+\pfour2143)\\
&\phantom{{}={}}-(8+2\epsilon_{3a}+2i\epsilon_{3b})(\pfour1243+\pfour1432)\\
&\phantom{{}={}}
-(8+2\epsilon_{3a}-2i\epsilon_{3b})(\pfour2134+\pfour3214)\\
&\phantom{{}={}}-10(\pfour1234+\pfour4321)
\col
\end{aligned}
\end{equation}
where $\epsilon_{3a}$, $\epsilon_{3b}$, $\epsilon_{3c}$, $\epsilon_{3d}$ 
parameterize the free choice of 
the renormalization scheme, and $\beta=4\zeta(3)$ comes from the dressing
phase. The permutation structures are defined as
\begin{equation}
\pthree{a_1}{\dots}{a_n}=\sum_{r=0}^{L-1}\perm_{a_1+r,\;a_1+r+1}\cdots\perm_{a_n+r,\;a_n+r+1}
\end{equation}
for the action on a cyclic state with $L$ sites, where $\perm_{a,\;a+1}$ 
permutes the flavours of the $a$-th and $(a+1)$-th site.
Some rules for the manipulation of these structures can be found in 
\cite{Beisert:2005wv}.

In order to obtain the four-loop contributions we are interested in, we cannot use the expression \eqref{D4} directly since it contains terms which
describe the permutations among five neighbouring legs. 
Hence it can
be applied only to a state in the asymptotic sense, 
i.e.\ the number of sites in the state
has to be five or more. If we want to obtain the sum of
all four-loop Feynman diagrams using $D_4$, we can correct it for the application  
on a length four state: the 
contributions from all the diagrams which describe the interactions of 
five neighbouring legs have to be replaced by the contributions from all 
four-loop wrapping interactions. 

The flavour permutation structure of each Feynman diagram is completely 
determined by the scalar interactions.
As will be explained in \cite{Fiamberti:2008} the 
relevant flavour exchanges 
can be uniquely captured in terms of the four functions
\begin{equation}\label{chistruc}
\begin{aligned}
\chi(a,b,c,d)&=\pid-4\pone1
+\ptwo ab+\ptwo ac+\ptwo ad+\ptwo bc+\ptwo bd+\ptwo cd\\
&\phantom{{}={}}
-\pthree abc-\pthree abd-\pthree acd-\pthree bcd
+\pfour abcd\col\\
\chi(a,b,c)&=-\pid+3\pone1
-\ptwo ab-\ptwo ac-\ptwo bc+\pthree abc\col\\
\chi(a,b)&=\pid-2\pone1+\ptwo ab\col\\
\chi(1)&=-\pid+\pone1
\col
\end{aligned}
\end{equation}
where the number of arguments $a,b,c,d=1,\dots,4$ is given by the
number of four-vertices. 
The independent flavour-exchange functions for the range five interactions are 
found by replacing $a,b,c,d$ with the corresponding arguments of the range 
five permutation structures found in \eqref{D4}. 

We have considered the contributions of all four-loop range five Feynman
diagrams. As one important result we find that those diagrams, in 
which  the first or the fifth line interacts with the rest of the graph 
only via flavour-neutral gauge bosons,
cancel  against each other. 
This means that the range five contributions can be extracted 
directly from $D_4$ given in \eqref{D4}. 
The terms we have to subtract from $D_4$ are 
uniquely given as a linear combination of the 
flavour exchange functions \eqref{chistruc} which cancels all the range five 
permutation structures in \eqref{D4}. 
They read
\begin{equation}
\begin{aligned}\label{r5D}
\delta D_4
&=-10[\chi(1,2,3,4)+\chi(4,3,2,1)]+(18+4\epsilon_{3a})[\chi(1,3,2,4)+\chi(2,1,4,3)]\\
&\phantom{{}={}}
-(8+2\epsilon_{3a}+2i\epsilon_{3b})[\chi(1,2,4,3)+\chi(1,4,3,2)]\\
&\phantom{{}={}}
-(8+2\epsilon_{3a}-2i\epsilon_{3b})[\chi(2,1,3,4)+\chi(3,2,1,4)]\\
&\phantom{{}={}}
-(4+4i\epsilon_{3b}+2i\epsilon_{3c})[\chi(1,2,4)+\chi(1,4,3)]\\
&\phantom{{}={}}
-(4-4i\epsilon_{3b}-2i\epsilon_{3c})[\chi(1,3,4)+\chi(2,1,4)]\\
&\phantom{{}={}}
-4\chi(1,4)
\pnt
\end{aligned}
\end{equation}
Here we stress that the subtraction is not only a simple subtraction of all
range five permutation structures as attempted in \cite{Fischbacher:2004iu} 
for the BMN matrix model. 
The above subtraction modifies also the coefficients of the 
permutation structures of lower range in \eqref{D4}.

We have checked the above constructed range five contributions 
\eqref{r5D} with an explicit Feynman graph computation. In the used MS scheme 
we find
\begin{equation}\label{epsilons}
\epsilon_{3a}=-4\col\qquad
\epsilon_{3b}=-i\frac{4}{3}\col\qquad
\epsilon_{3c}=i\frac{4}{3}\pnt
\end{equation}


Now we can apply the subtracted dilatation operator to the states with $L=4$ 
sites. 
In the $SU(2)$ subsector there exist two composite operators of length $L=4$
and with two `impurities'
which mix under renormalization. The corresponding states are given by
\begin{equation}\label{Opbasis}
\mathcal{O}_1=\tr(\phi Z\phi Z)\col\qquad\mathcal{O}_2=\tr(\phi\phi ZZ)\pnt
\end{equation}
Defining a two-dimensional vector 
$\vec{\mathcal{O}}=(\mathcal{O}_1,\mathcal{O}_2)^\text{t}$, the 
result for the subtracted dilatation operator becomes
\begin{equation}\label{D4submatrix}
D_4^\text{sub}\equiv D_4-\delta D_4\to
4\left(121+12\zeta(3)\right)M
\col
\end{equation}
where the mixing matrix is given by
\begin{equation}\label{mmatrix}
M=\begin{pmatrix} -4 & 4 \\ 2 & -2 \end{pmatrix}
\pnt
\end{equation}
Now to the subtracted dilatation operator we add the wrapping 
interactions.

\section{Wrapping interactions}

In \cite{Sieg:2005kd} a systematic Feynman-diagrammatic 
analysis of wrapping interactions has been performed.
It was shown that 
the wrapping interactions have a genus changing effect. This means in 
particular
that they are responsible for 
a non-trivial map  between the planar part of the 
dilatation operator and the 
planar part of the $2$-point functions of composite single trace operators. 
Also several unique properties of the wrapping interactions have been
worked out, and a systematic method for projecting out them
has been proposed. 
We want to make our calculation using $\mathcal{N}=1$ superspace
techniques, hence we will adapt this direct approach to construct all wrapping
supergraphs. 

The $\mathcal{N}=4$ SYM action written in terms of $\mathcal{N}=1$ 
superfields is given by (we use notations and conventions of 
\cite{Gates:1983nr})

\begin{equation}
\begin{aligned}
S&= \int \de^4x\de^4\theta
\tr \left(\e^{-g_\YM V} \bar \phi_i\e^{g_\YM V} 
\phi^i\right) 
+\frac{1}{2g_\YM^2}\int\de^4x\de^2\theta\tr\left(W^\alpha W_\alpha\right)\\
&\phantom{{}={}}+{\rm i}g_\YM\int\de^4x\de^2\theta
\tr\left(\phi_1\comm{\phi_2}{\phi_3}\right) +\text{h.c.}
\end{aligned}
\end{equation}
where $W_\alpha = {\rm i} \bar D^2 \left(e^{-g_\YM V} D_\alpha\,e^{g_\YM V}\right)$, and
$V=V^aT^a$, $\phi_i=\phi_i^aT^a$, {\small $i=1,2,3$}, $T^a$ being $SU(N)$
matrices. We will usually denote the three chiral superfields $\phi^i$ as 
$(\phi,\psi,Z)$, using the same letters used before for their lowest 
components.

Following \cite{Sieg:2005kd} we then proceed to construct all wrapping
supergraphs contributing to the renormalization of the chiral operators
\eqref{Opbasis}.
To do this, we rely on the following
findings, specialized to the length $L=4$ diagrams 
\begin{enumerate}
\item Wrapping diagrams which differ by an application of the cyclic rotations
$C_4\times C_4$, acting respectively on the incoming and outgoing four
legs, give the same contribution to the dilatation operator. They are just
different graphical representations of a single diagram in the two-dimensional
plane.
\item After removing the composite operator(s), the three-loop range four 
diagrams are simply-connected tree level graphs.
\item Wrapping diagrams containing at least one gauge boson line can be 
constructed by adding a single (wrapping) vector line to three-loop range four 
diagrams.
\item The remaining few fully chiral wrapping diagrams can be directly built 
by inspection of the Feynman rules.
\end{enumerate}
After constructing all diagrams of this type, we have to perform standard 
superspace $D$-algebra. This reduces the various contributions to ordinary 
massless four-loop momentum integrals. 
The anomalous dimension is then given in terms of the divergent part of these 
integrals. In a dimensional regularization approach it is given directly
by the $\frac{1}{\varepsilon}$ pole.

We list here only the results of the calculation, leaving a
detailed discussion for the forthcoming paper \cite{Fiamberti:2008}. 
First we give 
 a description of the needed integrals which have been computed using
dimensional regularization with $D=4-2\varepsilon$ in the MS scheme. 

After completion of the $D$-algebra we produce
both integrals without and with derivatives.
The first ones are computed with the 
method of uniqueness \cite{Kazakov:1983ns}.
We find for the overall poles, i.e.\ after subtraction of  subdivergencies
\settoheight{\eqoff}{$\times$}%
\setlength{\eqoff}{0.5\eqoff}%
\addtolength{\eqoff}{-7.5\unitlength}%
\begin{equation}
\begin{aligned}\label{scalarintegrals}
\raisebox{\eqoff}{%
\begin{fmfchar*}(20,15)
  \fmfleft{in}
  \fmfright{out}
  \fmf{plain}{in,v1}
  \fmf{plain,left=0.25}{v1,v2}
  \fmf{plain,left=0}{v2,v4}
  \fmf{plain,left=0.25}{v4,v3}
  \fmf{plain,tension=0.5,right=0.25}{v1,v0,v1}
  \fmf{plain,right=0.25}{v0,v3}
  \fmf{plain}{v0,v2}
  \fmf{plain}{v0,v4}
  \fmf{plain}{v3,out}
\fmffixed{(0.9w,0)}{v1,v3}
\fmffixed{(0.4w,0)}{v2,v4}
\fmfpoly{phantom}{v4,v2,v0}
\fmffreeze
\end{fmfchar*}}
&=\frac{1}{(4\pi)^8}\Big(
-\frac{1}{24\varepsilon^4}+\frac{1}{4\varepsilon^3}
-\frac{19}{24\varepsilon^2}
+\frac{5}{4\varepsilon}
\Big)\col
\\
\raisebox{\eqoff}{%
\begin{fmfchar*}(20,15)
  \fmfleft{in}
  \fmfright{out}
  \fmf{plain}{in,v1}
  \fmf{plain,left=0.25}{v1,v2}
  \fmf{plain,left=0.25}{v2,v3}
  \fmf{plain,left=0.25}{v3,v4}
  \fmf{plain,left=0.25}{v4,v1}
  \fmf{plain,tension=0.5,right=0.5}{v2,v0,v2}
  \fmf{phantom}{v0,v3}
  \fmf{plain}{v1,v0}
  \fmf{plain}{v0,v4}
  \fmf{plain}{v3,out}
\fmffixed{(0.9w,0)}{v1,v3}
\fmffixed{(0,0.45w)}{v4,v2}
\fmffreeze
\end{fmfchar*}}
&=\frac{1}{(4\pi)^8}\Big(
-\frac{1}{24\varepsilon^4}+\frac{1}{4\varepsilon^3}
-\frac{19}{24\varepsilon^2}
+\frac{1}{\varepsilon}\Big(\frac{5}{4}-\zeta(3)\Big)
\Big)\col
\\
\raisebox{\eqoff}{%
\begin{fmfchar*}(20,15)
  \fmfleft{in}
  \fmfright{out}
  \fmf{plain}{in,v1}
  \fmf{plain,left=0.25}{v1,v2}
  \fmf{plain,left=0.25}{v2,v3}
  \fmf{plain,left=0.25}{v3,v4}
  \fmf{plain,left=0.25}{v4,v1}
  \fmf{plain,tension=0.5,right=0.25}{v1,v0,v1}
  \fmf{phantom}{v0,v3}
  \fmf{plain}{v2,v0}
  \fmf{plain}{v0,v4}
  \fmf{plain}{v3,out}
\fmffixed{(0.9w,0)}{v1,v3}
\fmffixed{(0,0.45w)}{v4,v2}
\fmffreeze
\end{fmfchar*}}
&=\frac{1}{(4\pi)^8}\Big(
-\frac{1}{12\varepsilon^4}+\frac{1}{3\varepsilon^3}
-\frac{5}{12\varepsilon^2}
-\frac{1}{\varepsilon}\Big(\frac{1}{2}-\zeta(3)\Big)\Big)\col
\\
\raisebox{\eqoff}{%
\begin{fmfchar*}(20,15)
  \fmfleft{in}
  \fmfright{out}
  \fmf{plain}{in,v1}
  \fmf{plain,left=0.25}{v1,v2}
  \fmf{plain,left=0.25}{v2,v3}
  \fmf{plain,left=0.25}{v3,v4}
  \fmf{plain,left=0.25}{v4,v1}
  \fmf{plain,tension=0.5,right=0.5}{v2,v0,v2}
  \fmf{plain,tension=0.5,right=0.5}{v0,v4,v0}
  \fmf{plain}{v3,out}
\fmffixed{(0.9w,0)}{v1,v3}
\fmffixed{(0,0.45w)}{v4,v2}
\fmffreeze
\end{fmfchar*}}
&=\frac{1}{(4\pi)^8}\Big(
-\frac{1}{6\varepsilon^4}+\frac{1}{3\varepsilon^3}
+\frac{1}{3\varepsilon^2}
-\frac{1}{\varepsilon}(1-\zeta(3))
\Big)\pnt
\end{aligned}
\end{equation}

The integrals with derivatives (i.e.\ with non-trivial numerators) 
which cannot be rephrased in terms 
of integrals of the previous type, are calculated with the Gegenbauer 
polynomial $x$-space 
technique (GPXT) \cite{Chetyrkin:1980pr}, and are independently checked with 
the help of {\tt MINCER} \cite{Larin:1998}, a computer program to compute 
$3$-loop integrals.
Furthermore we have used the GPXT and {\tt MINCER} to check once again
the above results for the integrals without derivatives.

In particular the GPXT turns out to be very useful in 
the following situation:
\begin{enumerate}
\item The integral contains a vertex at which a large number of propagators 
end (in our case provided by the composite operator) which is chosen as the 
`root vertex'. 
\item The corresponding angular graph is planar, and it only contains 
loops which can be resolved by a sequential use of the Clebsch-Gordan series 
for the Gegenbauer polynomials.
\item The integral can be rearranged such that the linear combinations of 
momenta in the numerator 
become the momenta of propagators which end at the 
root vertex.
\item Only the pole structures of the integrals are of interest.
\end{enumerate}
With the GPXT we can then find analytic expressions for the pole structure 
of the required four-loop integrals. In particular, the fact that we are only 
interested in the pole structure allows us to considerably simplify the 
procedure by introducing an alternative IR regularization procedure into the
$x$-space integrals. 

We will describe the GPXT, and its modifications in detail in 
\cite{Fiamberti:2008}. 

The overall poles of the integrals 
with a non-trivial numerator, which are required to compute the wrapping 
contributions, are given by
\begin{equation}\label{derivativeintegrals}
\begin{gathered}
\begin{aligned}
\raisebox{\eqoff}{%
\begin{fmfchar*}(20,15)
  \fmfleft{in}
  \fmfright{out}
  \fmf{plain}{in,v1}
  \fmf{plain,left=0.25}{v1,v2}
  \fmf{plain,left=0.25}{v2,v3}
  \fmf{derplain,left=0.25}{v4,v1}
  \fmf{plain,right=0.25}{v4,v0}
  \fmf{plain,right=0.25}{v0,v5}
  \fmf{plain,right=0.75}{v4,v5}
  \fmf{derplain,right=0.25}{v5,v3}
  \fmf{plain}{v3,out}
\fmffixed{(0.9w,0)}{v1,v3}
\fmfpoly{phantom}{v2,v4,v5}
\fmffixed{(0.5w,0)}{v4,v5}
\fmffixed{(0.5w,0)}{v4,v5}
\fmf{plain,tension=0.25,right=0.25}{v2,v0,v2}
\fmffreeze
\fmfshift{(0,0.1w)}{in,out,v1,v2,v3,v4,v5,v0}
\end{fmfchar*}}
&=\frac{1}{(4\pi)^8}\Big(\frac{1}{12\varepsilon^2}
-\frac{7}{12\varepsilon}\Big)\col
\\
\raisebox{\eqoff}{%
\begin{fmfchar*}(20,15)
  \fmfleft{in}
  \fmfright{out}
  \fmf{plain}{in,v1}
  \fmf{plain,tension=2,left=0.125}{v1,v2c}
  \fmf{plain,tension=2,left=0.125}{v2c,v3}
  \fmf{plain,tension=1}{v2c,v2}
  \fmf{derplain,left=0.25}{v4,v1}
  \fmf{plain,right=0.25}{v4,v0}
  \fmf{plain,right=0}{v0,v1}
  \fmf{plain,right=0.25}{v0,v5}
  \fmf{plain,right=0.75}{v4,v5}
  \fmf{plain,right=0}{v3,v0}
  \fmf{derplain,right=0.25}{v5,v3}
  \fmf{phantom}{v3,out}
\fmffixed{(0,0.05w)}{v2c,v2}
\fmffixed{(0.9w,0)}{v1,v3}
\fmfpoly{phantom}{v2c,v4,v5}
\fmffixed{(0.5w,0)}{v4,v5}
\fmffreeze
\fmfshift{(0,0.15w)}{in,out,v1,v2,v2c,v3,v4,v5,v0}
\end{fmfchar*}}
&=
\frac{1}{(4\pi)^8}
\Big(\frac{1}{4\varepsilon^2}
-\frac{11}{12\varepsilon}\Big)\col
\end{aligned}
\qquad
\begin{aligned}
\raisebox{\eqoff}{%
\begin{fmfchar*}(20,15)
  \fmfleft{in}
  \fmfright{out}
  \fmf{plain}{in,v1}
  \fmf{plain,tension=2,left=0.25}{v1,v2}
  \fmf{plain,tension=2,left=0.25}{v2,v3}
  \fmf{derplain,left=0.25}{v4,v1}
  \fmf{plain,right=0.25}{v4,v0}
  \fmf{plain,right=0}{v0,v1}
  \fmf{plain,right=0.25}{v0,v5}
  \fmf{plain,right=0.75}{v4,v5}
  \fmf{phantom,right=0}{v3,v0}
  \fmf{derplain,right=0.25}{v5,v3}
  \fmf{plain}{v3,out}
\fmffixed{(0.9w,0)}{v1,v3}
\fmfpoly{phantom}{v2,v4,v5}
\fmffixed{(0.5w,0)}{v4,v5}
\fmf{plain,tension=0.5}{v2,v0}
\fmffreeze
\fmfshift{(0,0.15w)}{in,out,v1,v2,v3,v4,v5,v0}
\end{fmfchar*}}
&=
\frac{1}{(4\pi)^8}
\frac{1}{\varepsilon}(-\zeta(3))\col
\\
\raisebox{\eqoff}{%
\begin{fmfchar*}(20,15)
  \fmfleft{in}
  \fmfright{out}
  \fmf{plain}{in,v1}
  \fmf{derplain,tension=2,right=0.25}{v2,v1}
  \fmf{derplain,tension=2,left=0.25}{v2,v3}
  \fmf{plain,left=0.25}{v4,v1}
  \fmf{plain,right=0.25}{v4,v0}
  \fmf{plain,right=0}{v0,v1}
  \fmf{plain,right=0.25}{v0,v5}
  \fmf{plain,right=0.75}{v4,v5}
  \fmf{phantom,right=0}{v3,v0}
  \fmf{plain,right=0.25}{v5,v3}
  \fmf{plain}{v3,out}
\fmffixed{(0.9w,0)}{v1,v3}
\fmfpoly{phantom}{v2,v4,v5}
\fmffixed{(0.5w,0)}{v4,v5}
\fmf{plain,tension=0.5}{v2,v0}
\fmffreeze
\fmfshift{(0,0.15w)}{in,out,v1,v2,v3,v4,v5,v0}
\end{fmfchar*}}
&=\frac{1}{(4\pi)^8}\frac{1}{\varepsilon}\Big(\frac{1}{2}\zeta(3)
-\frac{5}{2}\zeta(5)\Big)\col
\end{aligned}
\\
\begin{aligned}
\raisebox{\eqoff}{%
\begin{fmfchar*}(20,15)
  \fmfleft{in}
  \fmfright{out}
  \fmf{plain}{in,v1}
  \fmf{derplain,tension=2,right=0.25}{v2,v1}
  \fmf{derplain,tension=2,left=0.25}{v3,v4}
  \fmf{derplainpt,tension=2,left=0.25}{v5,v1}
  \fmf{derplainpt,tension=2,right=0.25}{v6,v4}
  \fmf{plain}{v2,v0}
  \fmf{plain}{v3,v0}
  \fmf{plain}{v5,v0}
  \fmf{plain}{v6,v0}
  \fmf{plain}{v4,out}
\fmffixed{(0.9w,0)}{v1,v4}
\fmfpoly{phantom}{v3,v2,v5,v6}
  \fmf{plain}{v2,v3}
  \fmf{plain}{v5,v6}
\fmffixed{(0.4w,0)}{v5,v6}
\fmffreeze
\end{fmfchar*}}
&=
\frac{1}{(4\pi)^8}\frac{1}{\varepsilon}
\Big(-\frac{1}{2}-\frac{1}{2}\zeta(3)+\frac{5}{2}\zeta(5)\Big)\col
\\
\raisebox{\eqoff}{%
\begin{fmfchar*}(20,15)
  \fmfleft{in}
  \fmfright{out}
  \fmf{plain}{in,v1}
  \fmf{derplain,tension=2,right=0.25}{v2,v1}
  \fmf{derplain,tension=2,left=0.25}{v3,v4}
  \fmf{plain,tension=2,right=0.25}{v1,v5}
  \fmf{plain,tension=2,right=0.25}{v6,v4}
  \fmf{plain}{v2,v0}
  \fmf{plain}{v3,v0}
  \fmf{plain}{v5,v0}
  \fmf{plain}{v0,v6}
  \fmf{plain}{v4,out}
\fmffixed{(0.9w,0)}{v1,v4}
\fmfpoly{phantom}{v3,v2,v5,v6}
  \fmf{derplainpt}{v2,v3}
  \fmf{derplainpt}{v5,v6}
\fmffixed{(0.4w,0)}{v5,v6}
\fmffreeze
\end{fmfchar*}}
&=
\frac{1}{(4\pi)^8}\frac{1}{\varepsilon}
\Big(-\frac{1}{4}-\frac{3}{2}\zeta(3)
+\frac{5}{2}\zeta(5)\Big)\col
\\
\raisebox{\eqoff}{%
\begin{fmfchar*}(20,15)
  \fmfleft{in}
  \fmfright{out}
  \fmf{plain}{in,v1}
  \fmf{derplain,tension=2,right=0.25}{v2,v1}
  \fmf{plain,tension=2,left=0.25}{v3,v4}
  \fmf{plain,tension=2,right=0.25}{v1,v5}
  \fmf{derplainpt,tension=2,right=0.25}{v6,v4}
  \fmf{plain}{v2,v0}
  \fmf{plain}{v3,v0}
  \fmf{plain}{v5,v0}
  \fmf{plain}{v0,v6}
  \fmf{plain}{v4,out}
\fmffixed{(0.9w,0)}{v1,v4}
\fmfpoly{phantom}{v3,v2,v5,v6}
  \fmf{derplain}{v3,v2}
  \fmf{derplainpt}{v5,v6}
\fmffixed{(0.4w,0)}{v5,v6}
\fmffreeze
\end{fmfchar*}}
&=\frac{1}{(4\pi)^8}\frac{1}{\varepsilon}
\Big(-\frac{1}{8}-\frac{1}{4}\zeta(3)
+\frac{5}{4}\zeta(5)\Big)
\col
\end{aligned}
\end{gathered}
\end{equation}
where two arrows of the same type indicate a scalar product of the 
corresponding momenta in the numerator.

We now go on and list the supergraphs. Among the potentially contributing
diagrams a lot of cancellations already take place at intermediate steps
of the D-algebra. We give in Figure \ref{supergraphs} a list of the
supergraph structures with the corresponding integrals that are obtained
after D-algebra. The shown results contain also the factors coming from 
combinatorics and colour.
\begin{figure}[p]
\unitlength=0.75mm
\settoheight{\eqoff}{$\times$}%
\setlength{\eqoff}{0.5\eqoff}%
\addtolength{\eqoff}{-10\unitlength}%
\settoheight{\eqofftwo}{$\times$}%
\setlength{\eqofftwo}{0.5\eqofftwo}%
\addtolength{\eqofftwo}{-7.5\unitlength}%
\begin{equation*}\label{wrappingint}
\begin{aligned}
{}W_1&=\;
\raisebox{\eqoff}{%
\begin{fmfchar*}(20,20)
\fmftop{v1}
\fmfbottom{v5}
\fmfforce{(0.125w,h)}{v1}
\fmfforce{(0.125w,0)}{v5}
\fmffixed{(0.25w,0)}{v1,v2}
\fmffixed{(0.25w,0)}{v2,v3}
\fmffixed{(0.25w,0)}{v3,v4}
\fmffixed{(0.25w,0)}{v5,v6}
\fmffixed{(0.25w,0)}{v6,v7}
\fmffixed{(0.25w,0)}{v7,v8}
%
\fmf{plain,tension=0.5,right=0.25}{v2,vc1}
\fmf{plain,tension=0.5,left=0.25}{v3,vc1}
  \fmf{plain}{vc1,vc2}
\fmf{plain,tension=0.5,left=0.25}{v6,vc2}
\fmf{plain,tension=0.5,right=0.25}{v7,vc2}
\fmf{plain}{v5,v1}
\fmf{plain}{v8,v4}
\fmf{plain,tension=0.5,right=0,width=1mm}{v5,v8}
\fmffreeze
\end{fmfchar*}}
+ 3\;\raisebox{\eqoff}{%
\begin{fmfchar*}(10,20)
\fmfleft{v1}\fmfright{v2}
\fmf{photon}{v1,v2}
\end{fmfchar*}}\;
\to2\left(
\raisebox{\eqofftwo}{%
\begin{fmfchar*}(20,15)
  \fmfleft{in}
  \fmfright{out}
  \fmf{plain}{in,v1}
  \fmf{plain,left=0.25}{v1,v2}
  \fmf{plain,left=0}{v2,v4}
  \fmf{plain,left=0.25}{v4,v3}
  \fmf{plain,tension=0.5,right=0.25}{v1,v0,v1}
  \fmf{plain,right=0.25}{v0,v3}
  \fmf{plain}{v0,v2}
  \fmf{plain}{v0,v4}
  \fmf{plain}{v3,out}
\fmffixed{(0.9w,0)}{v1,v3}
\fmffixed{(0.4w,0)}{v2,v4}
\fmfpoly{phantom}{v4,v2,v0}
\fmffreeze
\end{fmfchar*}}
-
\raisebox{\eqofftwo}{%
\begin{fmfchar*}(20,15)
  \fmfleft{in}
  \fmfright{out}
  \fmf{plain}{in,v1}
  \fmf{plain,left=0.25}{v1,v2}
  \fmf{plain,left=0.25}{v2,v3}
  \fmf{plain,left=0.25}{v3,v4}
  \fmf{plain,left=0.25}{v4,v1}
  \fmf{plain,tension=0.5,right=0.5}{v2,v0,v2}
  \fmf{phantom}{v0,v3}
  \fmf{plain}{v1,v0}
  \fmf{plain}{v0,v4}
  \fmf{plain}{v3,out}
\fmffixed{(0.9w,0)}{v1,v3}
\fmffixed{(0,0.45w)}{v4,v2}
\fmffreeze
\end{fmfchar*}}
\right)
\chi(1)
=2\zeta(3)M\col
\\[0.5\baselineskip]
W_2&=\;
\raisebox{\eqoff}{%
\begin{fmfchar*}(20,20)
\fmftop{v1}
\fmfbottom{v5}
\fmfforce{(0.125w,h)}{v1}
\fmfforce{(0.125w,0)}{v5}
\fmffixed{(0.25w,0)}{v1,v2}
\fmffixed{(0.25w,0)}{v2,v3}
\fmffixed{(0.25w,0)}{v3,v4}
\fmffixed{(0.25w,0)}{v5,v6}
\fmffixed{(0.25w,0)}{v6,v7}
\fmffixed{(0.25w,0)}{v7,v8}
%
\fmf{plain,tension=0.5,right=0.25}{v1,vc1}
\fmf{plain,tension=0.5,left=0.25}{v2,vc1}
\fmf{plain,tension=0.5,right=0.25}{v3,vc2}
\fmf{plain,tension=0.5,left=0.25}{v4,vc2}
  \fmf{plain}{vc1,vc3}
  \fmf{plain}{vc2,vc4}
\fmf{plain,tension=0.5,left=0.25}{v5,vc3}
\fmf{plain,tension=0.5,right=0.25}{v6,vc3}
\fmf{plain,tension=0.5,left=0.25}{v7,vc4}
\fmf{plain,tension=0.5,right=0.25}{v8,vc4}
\fmf{plain,tension=0.5,right=0,width=1mm}{v5,v8}
\fmffreeze
\end{fmfchar*}}
+ 2\;\raisebox{\eqoff}{%
\begin{fmfchar*}(10,20)
\fmfleft{v1}\fmfright{v2}
\fmf{photon}{v1,v2}
\end{fmfchar*}}\;
\to2\left(
\raisebox{\eqofftwo}{%
\begin{fmfchar*}(20,15)
  \fmfleft{in}
  \fmfright{out}
  \fmf{plain}{in,v1}
  \fmf{derplain,tension=2,right=0.25}{v2,v1}
  \fmf{derplain,tension=2,left=0.25}{v3,v4}
  \fmf{derplainpt,tension=2,left=0.25}{v5,v1}
  \fmf{derplainpt,tension=2,right=0.25}{v6,v4}
  \fmf{plain}{v2,v0}
  \fmf{plain}{v3,v0}
  \fmf{plain}{v5,v0}
  \fmf{plain}{v6,v0}
  \fmf{plain}{v4,out}
\fmffixed{(0.9w,0)}{v1,v4}
\fmfpoly{phantom}{v3,v2,v5,v6}
  \fmf{plain}{v2,v3}
  \fmf{plain}{v5,v6}
\fmffixed{(0.4w,0)}{v5,v6}
\fmffreeze
\end{fmfchar*}}
+
\raisebox{\eqofftwo}{%
\begin{fmfchar*}(20,15)
  \fmfleft{in}
  \fmfright{out}
  \fmf{plain}{in,v1}
  \fmf{derplain,tension=2,right=0.25}{v2,v1}
  \fmf{derplain,tension=2,left=0.25}{v3,v4}
  \fmf{plain,tension=2,right=0.25}{v1,v5}
  \fmf{plain,tension=2,right=0.25}{v6,v4}
  \fmf{plain}{v2,v0}
  \fmf{plain}{v3,v0}
  \fmf{plain}{v5,v0}
  \fmf{plain}{v0,v6}
  \fmf{plain}{v4,out}
\fmffixed{(0.9w,0)}{v1,v4}
\fmfpoly{phantom}{v3,v2,v5,v6}
  \fmf{derplainpt}{v2,v3}
  \fmf{derplainpt}{v5,v6}
\fmffixed{(0.4w,0)}{v5,v6}
\fmffreeze
\end{fmfchar*}}
-2\;
\raisebox{\eqofftwo}{%
\begin{fmfchar*}(20,15)
  \fmfleft{in}
  \fmfright{out}
  \fmf{plain}{in,v1}
  \fmf{derplain,tension=2,right=0.25}{v2,v1}
  \fmf{plain,tension=2,left=0.25}{v3,v4}
  \fmf{plain,tension=2,right=0.25}{v1,v5}
  \fmf{derplainpt,tension=2,right=0.25}{v6,v4}
  \fmf{plain}{v2,v0}
  \fmf{plain}{v3,v0}
  \fmf{plain}{v5,v0}
  \fmf{plain}{v0,v6}
  \fmf{plain}{v4,out}
\fmffixed{(0.9w,0)}{v1,v4}
\fmfpoly{phantom}{v3,v2,v5,v6}
  \fmf{derplain}{v3,v2}
  \fmf{derplainpt}{v5,v6}
\fmffixed{(0.4w,0)}{v5,v6}
\fmffreeze
\end{fmfchar*}}
\right)
\chi(1,3)\\
&\phantom{{}={}\;
\raisebox{\eqoff}{%
\begin{fmfchar*}(20,20)
\end{fmfchar*}}
+ 2\;\raisebox{\eqoff}{%
\begin{fmfchar*}(10,20)
\end{fmfchar*}}\;}
=2(1+3\zeta(3)-5\zeta(5))M\col
\\[0.5\baselineskip]
W_3&=\;
\raisebox{\eqoff}{%
\begin{fmfchar*}(20,20)
\fmftop{v1}
\fmfbottom{v5}
\fmfforce{(0.125w,h)}{v1}
\fmfforce{(0.125w,0)}{v5}
\fmffixed{(0.25w,0)}{v1,v2}
\fmffixed{(0.25w,0)}{v2,v3}
\fmffixed{(0.25w,0)}{v3,v4}
\fmffixed{(0.25w,0)}{v5,v6}
\fmffixed{(0.25w,0)}{v6,v7}
\fmffixed{(0.25w,0)}{v7,v8}
\fmffixed{(0,whatever)}{vc1,vc2} 
\fmffixed{(0,whatever)}{vc3,vc4}
\fmf{plain,tension=0.5,right=0.25}{v2,vc1}
\fmf{plain,tension=0.5,left=0.25}{v3,vc1}
\fmf{plain,right=0.25}{v1,vc3}
\fmf{plain,tension=0.5,left=0.25}{v5,vc4}
\fmf{plain,tension=0.5,right=0.25}{v6,vc4}
\fmf{plain,right=0.25}{v7,vc2}
\fmf{plain}{v8,v4}
  \fmf{plain,tension=1}{vc1,vc2}
  \fmf{plain,tension=0.5}{vc2,vc3}
  \fmf{plain,tension=1}{vc3,vc4}
\fmf{plain,tension=0.5,right=0,width=1mm}{v5,v8}
\fmffreeze
\end{fmfchar*}}
+ 2\;\raisebox{\eqoff}{%
\begin{fmfchar*}(10,20)
\fmfleft{v1}\fmfright{v2}
\fmf{photon}{v1,v2}
\end{fmfchar*}}\;
\to4
\left(\raisebox{\eqofftwo}{%
\begin{fmfchar*}(20,15)
  \fmfleft{in}
  \fmfright{out}
  \fmf{plain}{in,v1}
  \fmf{plain,tension=2,left=0.25}{v1,v2}
  \fmf{plain,tension=2,left=0.25}{v2,v3}
  \fmf{derplain,left=0.25}{v4,v1}
  \fmf{plain,right=0.25}{v4,v0}
  \fmf{plain,right=0}{v0,v1}
  \fmf{plain,right=0.25}{v0,v5}
  \fmf{plain,right=0.75}{v4,v5}
  \fmf{phantom,right=0}{v3,v0}
  \fmf{derplain,right=0.25}{v5,v3}
  \fmf{plain}{v3,out}
\fmffixed{(0.9w,0)}{v1,v3}
\fmfpoly{phantom}{v2,v4,v5}
\fmffixed{(0.5w,0)}{v4,v5}
\fmf{plain,tension=0.5}{v2,v0}
\fmffreeze
\fmfshift{(0,0.15w)}{in,out,v1,v2,v3,v4,v5,v0}
\end{fmfchar*}}
-
\raisebox{\eqofftwo}{%
\begin{fmfchar*}(20,15)
  \fmfleft{in}
  \fmfright{out}
  \fmf{plain}{in,v1}
  \fmf{derplain,tension=2,right=0.25}{v2,v1}
  \fmf{derplain,tension=2,left=0.25}{v2,v3}
  \fmf{plain,right=0.25}{v1,v4}
  \fmf{plain,right=0.25}{v4,v0}
  \fmf{plain,right=0}{v0,v1}
  \fmf{plain,right=0.25}{v0,v5}
  \fmf{plain,right=0.75}{v4,v5}
  \fmf{phantom,right=0}{v3,v0}
  \fmf{plain,right=0.25}{v5,v3}
  \fmf{plain}{v3,out}
\fmffixed{(0.9w,0)}{v1,v3}
\fmfpoly{phantom}{v2,v4,v5}
\fmffixed{(0.5w,0)}{v4,v5}
\fmf{plain,tension=0.5}{v2,v0}
\fmffreeze
\fmfshift{(0,0.15w)}{in,out,v1,v2,v3,v4,v5,v0}
\end{fmfchar*}}\right)
\chi(1,2)
=2(3\zeta(3)-5\zeta(5))M\col
\\[0.5\baselineskip]
W_4&=\;
\raisebox{\eqoff}{%
\begin{fmfchar*}(20,20)
\fmftop{v1}
\fmfbottom{v5}
\fmfforce{(0.125w,h)}{v1}
\fmfforce{(0.125w,0)}{v5}
\fmffixed{(0.25w,0)}{v1,v2}
\fmffixed{(0.25w,0)}{v2,v3}
\fmffixed{(0.25w,0)}{v3,v4}
\fmffixed{(0.25w,0)}{v5,v6}
\fmffixed{(0.25w,0)}{v6,v7}
\fmffixed{(0.25w,0)}{v7,v8}
\fmffixed{(0,whatever)}{vc1,vc2} 
\fmffixed{(0,whatever)}{vc3,vc4}
\fmffixed{(0,whatever)}{vc5,vc6} 
\fmf{plain,tension=0.25,right=0.25}{v1,vc1}
\fmf{plain,tension=0.25,left=0.25}{v2,vc1}
\fmf{plain,left=0.25}{v5,vc2}
\fmf{plain,tension=1,left=0.25}{v3,vc3}
\fmf{plain,tension=1,left=0.25}{v4,vc5}
\fmf{plain,left=0.25}{v6,vc4}
\fmf{plain,tension=0.25,left=0.25}{v7,vc6}
\fmf{plain,tension=0.25,right=0.25}{v8,vc6}
  \fmf{plain,tension=0.5}{vc1,vc2}
  \fmf{plain,tension=0.5}{vc2,vc3}
  \fmf{plain,tension=0.5}{vc3,vc4}
  \fmf{plain,tension=0.5}{vc4,vc5}
  \fmf{plain,tension=0.5}{vc5,vc6}
\fmf{plain,tension=0.5,right=0,width=1mm}{v5,v8}
\fmffreeze
\end{fmfchar*}}
+1\;\raisebox{\eqoff}{%
\begin{fmfchar*}(10,20)
\fmfleft{v1}\fmfright{v2}
\fmf{photon}{v1,v2}
\end{fmfchar*}}\;
\to-4
\left(\raisebox{\eqofftwo}{%
\begin{fmfchar*}(20,15)
  \fmfleft{in}
  \fmfright{out}
  \fmf{plain}{in,v1}
  \fmf{plain,tension=2,left=0.25}{v1,v2}
  \fmf{plain,tension=2,left=0.25}{v2,v3}
  \fmf{derplain,left=0.25}{v4,v1}
  \fmf{plain,right=0.25}{v4,v0}
  \fmf{plain,right=0}{v0,v1}
  \fmf{plain,right=0.25}{v0,v5}
  \fmf{plain,right=0.75}{v4,v5}
  \fmf{phantom,right=0}{v3,v0}
  \fmf{derplain,right=0.25}{v5,v3}
  \fmf{plain}{v3,out}
\fmffixed{(0.9w,0)}{v1,v3}
\fmfpoly{phantom}{v2,v4,v5}
\fmffixed{(0.5w,0)}{v4,v5}
\fmf{plain,tension=0.5}{v2,v0}
\fmffreeze
\fmfshift{(0,0.15w)}{in,out,v1,v2,v3,v4,v5,v0}
\end{fmfchar*}}
-
\raisebox{\eqofftwo}{%
\begin{fmfchar*}(20,15)
  \fmfleft{in}
  \fmfright{out}
  \fmf{plain}{in,v1}
  \fmf{derplain,tension=2,right=0.25}{v2,v1}
  \fmf{derplain,tension=2,left=0.25}{v2,v3}
  \fmf{plain,right=0.25}{v1,v4}
  \fmf{plain,right=0.25}{v4,v0}
  \fmf{plain,right=0}{v0,v1}
  \fmf{plain,right=0.25}{v0,v5}
  \fmf{plain,right=0.75}{v4,v5}
  \fmf{phantom,right=0}{v3,v0}
  \fmf{plain,right=0.25}{v5,v3}
  \fmf{plain}{v3,out}
\fmffixed{(0.9w,0)}{v1,v3}
\fmfpoly{phantom}{v2,v4,v5}
\fmffixed{(0.5w,0)}{v4,v5}
\fmf{plain,tension=0.5}{v2,v0}
\fmffreeze
\fmfshift{(0,0.15w)}{in,out,v1,v2,v3,v4,v5,v0}
\end{fmfchar*}}\right)
\chi(1,2,3)
=2(3\zeta(3)-5\zeta(5))M\col
\\[0.5\baselineskip]
W_5&=\;
\raisebox{\eqoff}{%
\begin{fmfchar*}(20,20)
\fmftop{v1}
\fmfbottom{v5}
\fmfforce{(0.125w,h)}{v1}
\fmfforce{(0.125w,0)}{v5}
\fmffixed{(0.25w,0)}{v1,v2}
\fmffixed{(0.25w,0)}{v2,v3}
\fmffixed{(0.25w,0)}{v3,v4}
\fmffixed{(0.25w,0)}{v5,v6}
\fmffixed{(0.25w,0)}{v6,v7}
\fmffixed{(0.25w,0)}{v7,v8}
\fmffixed{(0,whatever)}{vc1,vc3} 
\fmffixed{(0,whatever)}{vc2,vc4}
\fmffixed{(0,whatever)}{vc5,vc6} 
\fmf{plain,tension=1,left=0.25}{v5,vc1}
\fmf{plain,tension=1,right=0.25}{v6,vc1}
\fmf{plain,tension=1,left=0.25}{v7,vc2}
\fmf{plain,tension=1,right=0.25}{v8,vc2}
\fmf{plain,tension=1,right=0.125}{v1,vc3}
\fmf{plain,tension=0.25,right=0.25}{v2,vc6}
\fmf{plain,tension=0.25,left=0.25}{v3,vc6}
\fmf{plain,tension=1,left=0.125}{v4,vc4}
  \fmf{plain,tension=4}{vc1,vc3}
  \fmf{plain,tension=4}{vc2,vc4}
  \fmf{plain,tension=0.5}{vc3,vc5}
  \fmf{plain,tension=0.5}{vc4,vc5}
  \fmf{plain,tension=1}{vc5,vc6}
\fmf{plain,tension=0.5,left=0,width=1mm}{v5,v8}
\fmffreeze
\end{fmfchar*}}
+ 1\;\raisebox{\eqoff}{%
\begin{fmfchar*}(10,20)
\fmfleft{v1}\fmfright{v2}
\fmf{photon}{v1,v2}
\end{fmfchar*}}\;
\to-2\;
\raisebox{\eqofftwo}{%
\begin{fmfchar*}(20,15)
  \fmfleft{in}
  \fmfright{out}
  \fmf{plain}{in,v1}
  \fmf{plain,tension=2,left=0.125}{v1,v2c}
  \fmf{plain,tension=2,left=0.125}{v2c,v3}
  \fmf{plain,tension=1}{v2c,v2}
  \fmf{derplain,left=0.25}{v4,v1}
  \fmf{plain,right=0.25}{v4,v0}
  \fmf{plain,right=0}{v0,v1}
  \fmf{plain,right=0.25}{v0,v5}
  \fmf{plain,right=0.75}{v4,v5}
  \fmf{plain,right=0}{v3,v0}
  \fmf{derplain,right=0.25}{v5,v3}
  \fmf{phantom}{v3,out}
\fmffixed{(0,0.05w)}{v2c,v2}
\fmffixed{(0.9w,0)}{v1,v3}
\fmfpoly{phantom}{v2c,v4,v5}
\fmffixed{(0.5w,0)}{v4,v5}
\fmffreeze
\fmfshift{(0,0.15w)}{in,out,v1,v2,v2c,v3,v4,v5,v0}
\end{fmfchar*}}
\;\chi(2,1,3)
=\frac{11}{3}M\col
\\[0.5\baselineskip]
W_6&=\;
\raisebox{\eqoff}{%
\begin{fmfchar*}(20,20)
\fmftop{v1}
\fmfbottom{v5}
\fmfforce{(0.125w,h)}{v1}
\fmfforce{(0.125w,0)}{v5}
\fmffixed{(0.25w,0)}{v1,v2}
\fmffixed{(0.25w,0)}{v2,v3}
\fmffixed{(0.25w,0)}{v3,v4}
\fmffixed{(0.25w,0)}{v5,v6}
\fmffixed{(0.25w,0)}{v6,v7}
\fmffixed{(0.25w,0)}{v7,v8}
\fmffixed{(0,whatever)}{vc1,vc3} 
\fmffixed{(0,whatever)}{vc2,vc4}
\fmffixed{(0,whatever)}{vc5,vc6} 
\fmf{plain,tension=1,right=0.25}{v1,vc1}
\fmf{plain,tension=1,left=0.25}{v2,vc1}
\fmf{plain,tension=1,right=0.25}{v3,vc2}
\fmf{plain,tension=1,left=0.25}{v4,vc2}
\fmf{plain,tension=1,left=0.125}{v5,vc3}
\fmf{plain,tension=0.25,left=0.25}{v6,vc6}
\fmf{plain,tension=0.25,right=0.25}{v7,vc6}
\fmf{plain,tension=1,right=0.125}{v8,vc4}
  \fmf{plain,tension=4}{vc1,vc3}
  \fmf{plain,tension=4}{vc2,vc4}
  \fmf{plain,tension=0.5}{vc3,vc5}
  \fmf{plain,tension=0.5}{vc4,vc5}
  \fmf{plain,tension=1}{vc5,vc6}
\fmf{plain,tension=0.5,right=0,width=1mm}{v5,v8}
\fmffreeze
\end{fmfchar*}}
+ 1\;\raisebox{\eqoff}{%
\begin{fmfchar*}(10,20)
\fmfleft{v1}\fmfright{v2}
\fmf{photon}{v1,v2}
\end{fmfchar*}}\;
\to-2\;
\raisebox{\eqofftwo}{%
\begin{fmfchar*}(20,15)
  \fmfleft{in}
  \fmfright{out}
  \fmf{plain}{in,v1}
  \fmf{plain,left=0.25}{v1,v2}
  \fmf{plain,left=0.25}{v2,v3}
  \fmf{derplain,left=0.25}{v4,v1}
  \fmf{plain,right=0.25}{v4,v0}
  \fmf{plain,right=0.25}{v0,v5}
  \fmf{plain,right=0.75}{v4,v5}
  \fmf{derplain,right=0.25}{v5,v3}
  \fmf{plain}{v3,out}
\fmffixed{(0.9w,0)}{v1,v3}
\fmfpoly{phantom}{v2,v4,v5}
\fmffixed{(0.5w,0)}{v4,v5}
\fmffixed{(0.5w,0)}{v4,v5}
\fmf{plain,tension=0.25,right=0.25}{v2,v0,v2}
\fmffreeze
\fmfshift{(0,0.1w)}{in,out,v1,v2,v3,v4,v5,v0}
\end{fmfchar*}}
\;\chi(1,3,2)
=\frac{7}{3}M\col
\\[0.5\baselineskip]
C_1&=\;
\raisebox{\eqoff}{%
\begin{fmfchar*}(20,20)
\fmftop{v1}
\fmfbottom{v5}
\fmfforce{(0.125w,h)}{v1}
\fmfforce{(0.125w,0)}{v5}
\fmffixed{(0.25w,0)}{v1,v2}
\fmffixed{(0.25w,0)}{v2,v3}
\fmffixed{(0.25w,0)}{v3,v4}
\fmffixed{(0.25w,0)}{v5,v6}
\fmffixed{(0.25w,0)}{v6,v7}
\fmffixed{(0.25w,0)}{v7,v8}
\fmffixed{(0,0.9w)}{v5,vh1}
\fmf{plain,tension=0.5,right=0.25}{v1,vc1}
\fmf{plain,tension=0.5,left=0.25}{v2,vc1}
\fmf{plain,tension=0.5,right=0.25}{v3,vc2}
\fmf{plain,tension=0.5,left=0.25}{v4,vc2}
  \fmf{plain}{vc1,vc3}
  \fmf{plain}{vc3,vc7}
  \fmf{plain}{vc7,vc5}
  \fmf{plain}{vc2,vc8}
  \fmf{plain}{vc8,vc4}
  \fmf{plain}{vc4,vc6}
  \fmf{plain,tension=0}{vc3,vc4}
\fmf{plain,tension=0.5,left=0.25}{v5,vc5}
\fmf{plain,tension=0.5,right=0.25}{v6,vc5}
\fmf{plain,tension=0.5,left=0.25}{v7,vc6}
\fmf{plain,tension=0.5,right=0.25}{v8,vc6}
\fmf{plain,tension=0.5,right=0,width=1mm}{v5,v8}
\fmffreeze
\fmfposition
\fmfipath{p[]}
\fmfiset{p1}{vloc(__vc7) ..controls (-0.125w,ypart(vloc(__vc7))) and (-0.125w,-0.1w) .. (xpart(vloc(__v5)),-0.1w)}
\fmfiset{p2}{(xpart(vloc(__v5)),-0.1w) ..(xpart(vloc(__v8)),-0.1w)}
\fmfiset{p3}{(xpart(vloc(__v8)),-0.1w) ..controls (1.125w,-0.1w) and (1.125w,ypart(vloc(__vc8))) .. vloc(__vc8)}
\fmfi{plain}{p1 ..p2 ..p3}
\end{fmfchar*}}\;
\to-2\;
\raisebox{\eqofftwo}{%
\begin{fmfchar*}(20,15)
  \fmfleft{in}
  \fmfright{out}
  \fmf{plain}{in,v1}
  \fmf{plain,left=0.25}{v1,v2}
  \fmf{plain,left=0.25}{v2,v3}
  \fmf{plain,left=0.25}{v3,v4}
  \fmf{plain,left=0.25}{v4,v1}
  \fmf{plain,tension=0.5,right=0.5}{v2,v0,v2}
  \fmf{plain,tension=0.5,right=0.5}{v0,v4,v0}
  \fmf{plain}{v3,out}
\fmffixed{(0.9w,0)}{v1,v3}
\fmffixed{(0,0.45w)}{v4,v2}
\fmffreeze
\end{fmfchar*}}
\;M=-2(\zeta(3)-1)M\col
\\[0.5\baselineskip]
C_2&=\;
\raisebox{\eqoff}{%
\begin{fmfchar*}(20,20)
\fmftop{v1}
\fmfbottom{v5}
\fmfforce{(0.125w,h)}{v1}
\fmfforce{(0.125w,0)}{v5}
\fmffixed{(0.25w,0)}{v1,v2}
\fmffixed{(0.25w,0)}{v2,v3}
\fmffixed{(0.25w,0)}{v3,v4}
\fmffixed{(0.25w,0)}{v5,v6}
\fmffixed{(0.25w,0)}{v6,v7}
\fmffixed{(0.25w,0)}{v7,v8}
\fmffixed{(whatever,0)}{vc1,vc3}
\fmffixed{(whatever,0)}{vc5,vc7}
\fmffixed{(whatever,0)}{vc3,vc4}
\fmffixed{(whatever,0)}{vc7,vc8}
\fmf{plain,tension=1,right=0.125}{v1,vc1}
\fmf{plain,tension=0.5,right=0.25}{v2,vc2}
\fmf{plain,tension=0.5,left=0.25}{v3,vc2}
\fmf{plain,tension=1,left=0.125}{v4,vc4}
\fmf{plain,tension=1,left=0.125}{v5,vc5}
\fmf{plain,tension=0.5,left=0.25}{v6,vc6}
\fmf{plain,tension=0.5,right=0.25}{v7,vc6}
\fmf{plain,tension=1,right=0.125}{v8,vc8}
\fmf{plain}{vc1,vc5}
\fmf{plain}{vc4,vc8}
\fmf{plain}{vc2,vc3}
\fmf{plain}{vc6,vc7}
\fmf{plain,tension=3}{vc3,vc7}
\fmf{plain,tension=0.5}{vc1,vc3}
\fmf{plain,tension=0.5}{vc7,vc8}
\fmf{phantom,tension=0.5}{vc5,vc7}
\fmf{phantom,tension=0.5}{vc3,vc4}
\fmf{plain,tension=0.5,right=0,width=1mm}{v5,v8}
\fmffreeze
\fmfposition
\fmfipath{p[]}
\fmfiset{p1}{vloc(__vc5) ..controls (-0.125w,ypart(vloc(__vc5))) and (-0.125w,-0.1w) .. (xpart(vloc(__v5)),-0.1w)}
\fmfiset{p2}{(xpart(vloc(__v5)),-0.1w) ..(xpart(vloc(__v8)),-0.1w)}
\fmfiset{p3}{(xpart(vloc(__v8)),-0.1w) ..controls (1.125w,-0.1w) and (1.125w,ypart(vloc(__vc4))) .. vloc(__vc4)}
\fmfi{plain}{p1 ..p2 ..p3}
\fmf{plain,tension=0.5,right=0,width=1mm}{v5,v8}
\fmffreeze
\end{fmfchar*}}\;
\to-2\;
\raisebox{\eqofftwo}{%
\begin{fmfchar*}(20,15)
  \fmfleft{in}
  \fmfright{out}
  \fmf{plain}{in,v1}
  \fmf{plain,left=0.25}{v1,v2}
  \fmf{plain,left=0.25}{v2,v3}
  \fmf{plain,left=0.25}{v3,v4}
  \fmf{plain,left=0.25}{v4,v1}
  \fmf{plain,tension=0.5,right=0.5}{v2,v0,v2}
  \fmf{phantom}{v0,v3}
  \fmf{plain}{v1,v0}
  \fmf{plain}{v0,v4}
  \fmf{plain}{v3,out}
\fmffixed{(0.9w,0)}{v1,v3}
\fmffixed{(0,0.45w)}{v4,v2}
\fmffreeze
\end{fmfchar*}}
\;M=-2\Big(\frac{5}{4}-\zeta(3)\Big)M\col
\\[0.5\baselineskip]
C_3&=\;
\raisebox{\eqoff}{%
\begin{fmfchar*}(20,20)
\fmftop{v1}
\fmfbottom{v5}
\fmfforce{(0.125w,h)}{v1}
\fmfforce{(0.125w,0)}{v5}
\fmffixed{(0.25w,0)}{v1,v2}
\fmffixed{(0.25w,0)}{v2,v3}
\fmffixed{(0.25w,0)}{v3,v4}
\fmffixed{(0.25w,0)}{v5,v6}
\fmffixed{(0.25w,0)}{v6,v7}
\fmffixed{(0.25w,0)}{v7,v8}
\fmffixed{(0,whatever)}{vc1,vc5} 
\fmffixed{(0,whatever)}{vc2,vc3}
\fmffixed{(0,whatever)}{vc3,vc6}
\fmffixed{(0,whatever)}{vc6,vc7}
\fmffixed{(0,whatever)}{vc4,vc8}
\fmffixed{(0.5w,0)}{vc1,vc4}
\fmffixed{(0.5w,0)}{vc5,vc8} 
\fmf{plain,tension=1,right=0.125}{v1,vc1}
\fmf{plain,tension=0.25,right=0.25}{v2,vc2}
\fmf{plain,tension=0.25,left=0.25}{v3,vc2}
\fmf{plain,tension=1,left=0.125}{v4,vc4}

\fmf{plain,tension=1,left=0.125}{v5,vc5}
\fmf{plain,tension=0.25,left=0.25}{v6,vc6}
\fmf{plain,tension=0.25,right=0.25}{v7,vc6}
\fmf{plain,tension=1,right=0.125}{v8,vc8}
  \fmf{plain,tension=0.5}{vc1,vc3}
  \fmf{plain,tension=0.5}{vc2,vc3}
  \fmf{plain,tension=0.5}{vc3,vc4}
  \fmf{plain,tension=0.5}{vc5,vc7}
  \fmf{plain,tension=0.5}{vc6,vc7}
  \fmf{plain,tension=0.5}{vc7,vc8}
  \fmf{plain,tension=2}{vc1,vc8}
  \fmf{phantom,tension=2}{vc5,vc4}
\fmffreeze
\fmfposition
\fmfipath{p[]}
\fmfiset{p1}{vloc(__vc5) ..controls (-0.125w,ypart(vloc(__vc5))) and (-0.125w,-0.1w) .. (xpart(vloc(__v5)),-0.1w)}
\fmfiset{p2}{(xpart(vloc(__v5)),-0.1w) ..(xpart(vloc(__v8)),-0.1w)}
\fmfiset{p3}{(xpart(vloc(__v8)),-0.1w) ..controls (1.125w,-0.1w) and (1.125w,ypart(vloc(__vc4))) .. vloc(__vc4)}
\fmfi{plain}{p1 ..p2 ..p3}
\fmf{plain,tension=1,left=0,width=1mm}{v5,v8}
\fmffreeze
\end{fmfchar*}}\;
\to-2\;
\raisebox{\eqofftwo}{%
\begin{fmfchar*}(20,15)
  \fmfleft{in}
  \fmfright{out}
  \fmf{plain}{in,v1}
  \fmf{plain,left=0.25}{v1,v2}
  \fmf{plain,left=0.25}{v2,v3}
  \fmf{plain,left=0.25}{v3,v4}
  \fmf{plain,left=0.25}{v4,v1}
  \fmf{plain,tension=0.5,right=0.25}{v1,v0,v1}
  \fmf{phantom}{v0,v3}
  \fmf{plain}{v2,v0}
  \fmf{plain}{v0,v4}
  \fmf{plain}{v3,out}
\fmffixed{(0.9w,0)}{v1,v3}
\fmffixed{(0,0.45w)}{v4,v2}
\fmffreeze
\end{fmfchar*}}
\;M=-2\Big(\zeta(3)-\frac{1}{2}\Big)M\pnt
\end{aligned}
\end{equation*}
\caption{Classes of four-loop wrapping graphs and their quantitative contributions. For the classes $W_i$, $i=1,\dots,6$ only the underlying chiral structure is shown. It has to be completed to the wrapping graphs 
by appropriately adding the indicated number of gauge boson lines. For the 
chiral classes $C_j$, $j=1,\dots3$ all interactions are present.  
}
\label{supergraphs}
\end{figure}
\clearpage
We have distinguished purely chiral wrapping graphs (denoted by $C$)
from the graphs $W$, which contain a number of gauge boson propagators
(indicated by the number in front of the wiggled line) which have to
be added to the corresponding chiral structure, such that the corresponding 
graph becomes a four-loop wrapping diagram. The complete list of all 
contributing diagrams will be given in \cite{Fiamberti:2008}.
The equalities in Figure \ref{supergraphs} present the explicit results
in the $L=4$ basis 
\eqref{Opbasis} after inserting the values for the integrals.

Thus we find that the sum of all wrapping terms contributes to $D_4$ 
as
\begin{equation}\label{D4wrapping}
D_4^\text{w}\to-8\Big(\frac{17}{2}+18\zeta(3)-30\zeta(5)\Big)M
\pnt
\end{equation}

\section{The final result}

We now collect our results from the subtracted
dilatation operator \eqref{D4submatrix} and from the wrapping part 
\eqref{D4wrapping}. The total contribution is given by

\begin{equation}\label{D4matrix}
D_4^\text{sub}+D_4^\text{w}
\to\big(416-96\zeta(3)+240\zeta(5)
\big)M
\pnt
\end{equation}
The non-vanishing eigenvalue of this matrix is
\begin{equation}
\gamma_4=-2496+576\zeta(3)-1440\zeta(5)
\pnt
\end{equation}
Restoring the dependence on the coupling constant \eqref{gdef},
and including also the contributions at lower orders \cite{Beisert:2006ez},
our final result for the planar anomalous dimension of the length four 
Konishi-descendant up to four loops reads
\begin{equation}
\gamma=4+12g^2-48g^4+336g^6+g^8(-2496+576\zeta(3)-1440\zeta(5))
\pnt
\end{equation}


We conclude with some comments. 
In our calculation we could partially save the knowledge of the asymptotic
dilatation operator
at four loops by suitably subtracting all range five contributions. 
The task was simplified by the cancellation of those range five Feynman graphs
with the first or the last line interacting with the rest of the graph 
only via flavour-neutral gauge bosons. This makes the explicit evaluation of 
all range five Feynman graphs not necessary.
We believe that the absence of these contributions persists also to
higher orders. This would be very important to check, since it would allow 
us to directly determine the necessary subtractions in the case of $D_K$ 
when applied to a length $L=K$ operator. 

The use of $\mathcal{N}=1$ supergraph techniques was very powerful
for the explicit Feynman graph evaluation of the wrapping part of the
calculation. We found the cancellation of the overwhelming majority of the 
potentially contributing supergraphs.




In the literature two different results for the Konishi anomalous dimension 
are conjectured on the basis of the Hubbard model \cite{Rej:2005qt} and on
an analysis of the BFKL equation \cite{Kotikov:2007cy}.
Our result obtained from explicit calculation differs from them.
In particular, the presence of a term proportional to 
$\zeta(5)$ is new. 
The compatibility of this term with transcendentality principles
deserves further investigation.

\subsubsection*{Note added in proof}

We have corrected a factor of $2$ missing in $W_5$ of Figure \ref{supergraphs} after the appearance of \cite{Bajnok:2008bm}.

\section*{Acknowledgements}

This work has been supported in part by INFN, PRIN prot.2005024045-002 and
the European Commission RTN program MRTN-CT-2004-005104.

\end{fmffile}

\footnotesize
\bibliographystyle{JHEP}
\bibliography{references}

\end{document}

\begin{equation}
\begin{aligned}
%
%
%
%
%
\raisebox{\eqoff}{%
\begin{fmfchar*}(20,15)
  \fmfleft{in}
  \fmfright{out}
  \fmf{plain}{in,v1}
  \fmf{plain,tension=2,left=0.25}{v1,v2}
  \fmf{plain,tension=2,left=0.25}{v2,v3}
  \fmf{derplain,right=0.25}{v1,v4}
  \fmf{plain,right=0.25}{v4,v0}
  \fmf{plain,right=0}{v0,v1}
  \fmf{plain,right=0.25}{v0,v5}
  \fmf{plain,right=0.75}{v4,v5}
  \fmf{phantom,right=0}{v3,v0}
  \fmf{derplain,right=0.25}{v5,v3}
  \fmf{plain}{v3,out}
\fmffixed{(0.9w,0)}{v1,v3}
\fmfpoly{phantom}{v2,v4,v5}
\fmffixed{(0.5w,0)}{v4,v5}
\fmf{plain,tension=0.5}{v2,v0}
\fmffreeze
\fmfshift{(0,0.15w)}{in,out,v1,v2,v3,v4,v5,v0}
\end{fmfchar*}}
-
\raisebox{\eqoff}{%
\begin{fmfchar*}(20,15)
  \fmfleft{in}
  \fmfright{out}
  \fmf{plain}{in,v1}
  \fmf{derplain,tension=2,left=0.25}{v1,v2}
  \fmf{derplain,tension=2,left=0.25}{v2,v3}
  \fmf{plain,right=0.25}{v1,v4}
  \fmf{plain,right=0.25}{v4,v0}
  \fmf{plain,right=0}{v0,v1}
  \fmf{plain,right=0.25}{v0,v5}
  \fmf{plain,right=0.75}{v4,v5}
  \fmf{phantom,right=0}{v3,v0}
  \fmf{plain,right=0.25}{v5,v3}
  \fmf{plain}{v3,out}
\fmffixed{(0.9w,0)}{v1,v3}
\fmfpoly{phantom}{v2,v4,v5}
\fmffixed{(0.5w,0)}{v4,v5}
\fmf{plain,tension=0.5}{v2,v0}
\fmffreeze
\fmfshift{(0,0.15w)}{in,out,v1,v2,v3,v4,v5,v0}
\end{fmfchar*}}
&=\frac{1}{(4\pi)^8}\frac{1}{\varepsilon}\Big(-\frac{3}{2}\zeta(3)
+\frac{5}{2}\zeta(5)\Big)
\\
\raisebox{\eqoff}{%
\begin{fmfchar*}(20,15)
  \fmfleft{in}
  \fmfright{out}
  \fmf{plain}{in,v1}
  \fmf{derplain,tension=2,left=0.25}{v1,v2}
  \fmf{derplain,tension=2,left=0.25}{v3,v4}
  \fmf{derplainpt,tension=2,right=0.25}{v1,v5}
  \fmf{derplainpt,tension=2,right=0.25}{v6,v4}
  \fmf{plain}{v2,v0}
  \fmf{plain}{v3,v0}
  \fmf{plain}{v5,v0}
  \fmf{plain}{v6,v0}
  \fmf{plain}{v4,out}
\fmffixed{(0.9w,0)}{v1,v4}
\fmfpoly{phantom}{v3,v2,v5,v6}
  \fmf{plain}{v2,v3}
  \fmf{plain}{v5,v6}
\fmffixed{(0.4w,0)}{v5,v6}
\fmffreeze
\end{fmfchar*}}
+
\raisebox{\eqoff}{%
\begin{fmfchar*}(20,15)
  \fmfleft{in}
  \fmfright{out}
  \fmf{plain}{in,v1}
  \fmf{derplain,tension=2,left=0.25}{v1,v2}
  \fmf{derplain,tension=2,left=0.25}{v3,v4}
  \fmf{plain,tension=2,right=0.25}{v1,v5}
  \fmf{plain,tension=2,right=0.25}{v6,v4}
  \fmf{plain}{v2,v0}
  \fmf{plain}{v3,v0}
  \fmf{plain}{v5,v0}
  \fmf{plain}{v0,v6}
  \fmf{plain}{v4,out}
\fmffixed{(0.9w,0)}{v1,v4}
\fmfpoly{phantom}{v3,v2,v5,v6}
  \fmf{derplainpt}{v2,v3}
  \fmf{derplainpt}{v5,v6}
\fmffixed{(0.4w,0)}{v5,v6}
\fmffreeze
\end{fmfchar*}}
-2\;
\raisebox{\eqoff}{%
\begin{fmfchar*}(20,15)
  \fmfleft{in}
  \fmfright{out}
  \fmf{plain}{in,v1}
  \fmf{derplain,tension=2,left=0.25}{v1,v2}
  \fmf{plain,tension=2,left=0.25}{v3,v4}
  \fmf{plain,tension=2,right=0.25}{v1,v5}
  \fmf{derplainpt,tension=2,right=0.25}{v6,v4}
  \fmf{plain}{v2,v0}
  \fmf{plain}{v3,v0}
  \fmf{plain}{v5,v0}
  \fmf{plain}{v0,v6}
  \fmf{plain}{v4,out}
\fmffixed{(0.9w,0)}{v1,v4}
\fmfpoly{phantom}{v3,v2,v5,v6}
  \fmf{derplain}{v2,v3}
  \fmf{derplainpt}{v5,v6}
\fmffixed{(0.4w,0)}{v5,v6}
\fmffreeze
\end{fmfchar*}}
&=
\frac{1}{(4\pi)^8}\frac{1}{2\varepsilon}\Big(
-1-3\zeta(3)+5\zeta(5)\Big)
\\
%
%
%
%
%
\raisebox{\eqoff}{%
\begin{fmfchar*}(20,15)
  \fmfleft{in}
  \fmfright{out}
  \fmf{plain}{in,v1}
  \fmf{phantom,tension=2,left=0.25}{v1,v2}
  \fmf{plain,tension=2,left=0.25}{v2,v3}
  \fmf{derplain,right=0.25}{v1,v4}
  \fmf{plain,right=0.25}{v4,v0}
  \fmf{plain,right=0}{v0,v1}
  \fmf{plain,right=0.25}{v0,v5}
  \fmf{plain,right=0.75}{v4,v5}
  \fmf{phantom,right=0}{v3,v0}
  \fmf{derplain,right=0.25}{v5,v3}
  \fmf{plain}{v3,out}
\fmffixed{(0.9w,0)}{v1,v3}
\fmfpoly{phantom}{v2,v4,v5}
\fmffixed{(0.5w,0)}{v4,v5}
\fmf{plain,tension=0.25,right=0.25}{v2,v0}
\fmf{plain,tension=0.25,right=0.25}{v0,v2}
\fmffreeze
\fmfshift{(0,0.15w)}{in,out,v1,v2,v3,v4,v5,v0}
\end{fmfchar*}}
&=\frac{1}{(4\pi)^8}
\Big(\frac{1}{12\varepsilon^2}
-\frac{1}{12\varepsilon}\Big)
\\
\raisebox{\eqoff}{%
\begin{fmfchar*}(20,15)
  \fmfleft{in}
  \fmfright{out}
  \fmf{plain}{in,v1}
  \fmf{plain,right=0.25}{v1,v4}
  \fmf{plain,right=0.25}{v4,v0}
  \fmf{derplain,right=0.25}{v0,v1}
  \fmf{plain,right=0.25}{v0,v5}
  \fmf{plain,right=0.25}{v4,v6}
  \fmf{plain,left=0.25}{v5,v6}
  \fmf{derplain,right=0.25}{v3,v0}
  \fmf{plain,right=0.25}{v5,v3}
  \fmf{plain}{v3,out}
\fmffixed{(0.9w,0)}{v1,v3}
\fmfpoly{phantom}{v0,v4,v6,v5}
\fmffixed{(0.45w,0)}{v4,v5}
\fmf{plain}{v6,v0}
\fmffreeze
\fmfshift{(0,0.15w)}{in,out,v1,v2,v3,v4,v5,v6,v0}
\end{fmfchar*}}
&=\frac{1}{(4\pi)^8}
\Big(-\frac{1}{4\varepsilon}\Big)
\\
\raisebox{\eqoff}{%
\begin{fmfchar*}(20,15)
  \fmfleft{in}
  \fmfright{out}
  \fmf{plain}{in,v1}
  \fmf{plain,right=0.25}{v1,v4}
  \fmf{plain,right=0.25}{v4,v0}
  \fmf{plain,right=0.25}{v0,v1}
  \fmf{plain,right=0.25}{v0,v5}
  \fmf{derplain,right=0.25}{v4,v6}
  \fmf{plain,left=0.25}{v5,v6}
  \fmf{plain,right=0.25}{v3,v0}
  \fmf{derplain,right=0.25}{v5,v3}
  \fmf{plain}{v3,out}
\fmffixed{(0.9w,0)}{v1,v3}
\fmfpoly{phantom}{v0,v4,v6,v5}
\fmffixed{(0.45w,0)}{v4,v5}
\fmf{plain}{v6,v0}
\fmffreeze
\fmfshift{(0,0.15w)}{in,out,v1,v2,v3,v4,v5,v6,v0}
\end{fmfchar*}}
&=\frac{1}{(4\pi)^8}
\Big(\frac{1}{4\varepsilon^2}
-\frac{5}{12\varepsilon}\Big)
\end{aligned}
\end{equation}

\settoheight{\eqoff}{$\times$}%
\setlength{\eqoff}{0.5\eqoff}%
\addtolength{\eqoff}{-10\unitlength}%
\begin{equation}
\begin{aligned}
\raisebox{\eqoff}{%
\begin{fmfchar*}(20,20)
\fmftop{v1}
\fmfbottom{v5}
\fmfforce{(0.125w,h)}{v1}
\fmfforce{(0.125w,0)}{v5}
\fmffixed{(0.25w,0)}{v1,v2}
\fmffixed{(0.25w,0)}{v2,v3}
\fmffixed{(0.25w,0)}{v3,v4}
\fmffixed{(0.25w,0)}{v5,v6}
\fmffixed{(0.25w,0)}{v6,v7}
\fmffixed{(0.25w,0)}{v7,v8}
%
\fmf{plain,tension=0.5,right=0.25}{v1,vc1}
\fmf{plain,tension=0.5,left=0.25}{v2,vc1}
\fmf{plain,tension=0.5,right=0.25}{v3,vc2}
\fmf{plain,tension=0.5,left=0.25}{v4,vc2}
  \fmf{plain}{vc1,vc3}
  \fmf{plain}{vc2,vc4}
\fmf{plain,tension=0.5,left=0.25}{v5,vc3}
\fmf{plain,tension=0.5,right=0.25}{v6,vc3}
\fmf{plain,tension=0.5,left=0.25}{v7,vc4}
\fmf{plain,tension=0.5,right=0.25}{v8,vc4}
\fmf{plain,tension=0.5,right=0,width=1mm}{v5,v8}
\fmffreeze
\fmfposition
\fmfipath{p[]}
\fmfipair{g[]}
\fmfiset{p5}{vpath(__v5,__vc3)}
\fmfiset{p6}{vpath(__v6,__vc3)}
\fmfiset{p7}{vpath(__v7,__vc4)}
\fmfiset{p8}{vpath(__v8,__vc4)}
\fmfiset{p9}{vpath(__v5,__v8)}
\fmfiset{g1}{point length(p5)/2 of p5}
\fmfiset{g2}{point length(p6)/2 of p6}
\fmfiset{g3}{point length(p7)/2 of p7}
\fmfiset{g4}{point length(p8)/2 of p8}
\fmfiset{g5}{point length(p9)/2 of p9}

\end{fmfchar*}}
\end{aligned}
\end{equation}
